\documentclass[]{aa}
\usepackage{psfig}
\usepackage{times}

\begin{document}
\newcommand{\pdrv}[2]{\frac{\partial #1}{\partial #2}}
\newcommand{\drv}[2]{{{{\rm d} #1}\over {{\rm d} #2}}}

   \thesaurus{06 	    
              (08.02.3;     
               08.05.3;     
               08.13.2;     
               08.14.1;     
               08.23.1;     
               03.13.4)}    
 
   \title{Formation of the binary pulsars PSR~B2303+46 and PSR~J1141--6545}
   \subtitle{-- young neutron stars with old white dwarf companions}

   \author{Thomas M. Tauris \inst{1} \& Thomas Sennels \inst{2}}

   \offprints{tauris@astro.uva.nl}

   \institute{Center for High-Energy Astrophysics,
              University of Amsterdam,
              Kruislaan 403, NL-1098 SJ Amsterdam, The Netherlands
              \and Institute of Physics \& Astronomy,
              Aarhus University,
              DK-8000 C, Aarhus, Denmark}

     \date{Received 10 September 1999 / Accepted 21 December 1999}

   \maketitle \markboth{T.M. Tauris \& T. Sennels: Formation of
                        PSR B2303+46 and PSR~J1141--6545}{}

\begin{abstract} 
We have investigated the formation of the binary radio pulsars
PSR~B2303+46 and PSR~J1141--6545 via Monte Carlo simulations of a large number of
interacting stars in binary systems. PSR~B2303+46 has recently
been shown (van~Kerkwijk \& Kulkarni 1999) to be the first neutron star
-- white dwarf binary system observed, in which the neutron star
was born {\em after} the formation of the white dwarf.
We discuss the formation process for such a system and are
able to put constraints on the parameters
of the initial ZAMS binary.\\
We present statistical evidence in favor
of a white dwarf companion to the binary pulsar PSR~J1141--6545,
just recently discovered in the Parkes Multibeam Survey.
If this is confirmed by observations
this system will be the second one known
in which the neutron star was born after its white dwarf companion.
We also predict a minimum space velocity of 150~km~s$^{-1}$
for PSR~J1141--6545, and show it must have experienced
an asymmetric SN in order to explain its low eccentricity.\\
Finally, we estimate the birthrate of these systems relative
to other binary pulsar systems and present the
expected distribution of their orbital periods, eccentricities
and velocities.

\end{abstract}

\section{Introduction}
Recent observations by van~Kerkwijk \& Kulkarni (1999) give evidence 
for a binary (PSR~B2303+46) in which the white dwarf was formed
before the neutron star (hereafter a WDNS system), as opposed 
to the normal case 
in which the neutron star is formed first (NSWD).
There are at present $\sim 40$ systems known in the Galactic disk
of the latter case. These are the millisecond pulsar binaries
in which an old neutron star was "recycled" via accretion
of angular momentum and mass from the giant progenitor of the
white dwarf (e.g. Alpar~et~al. 1982; van~den~Heuvel 1984). 
The third type of compact binaries in which a pulsar has
been detected, is the double neutron star systems (NSNS).
There are at present $\sim 5$ of these systems known 
(Nice~et~al. 1996; Manchester~et~al. 2000).
It is the formation of WDNS systems which is of interest
in this paper.
For a review on the formation and evolution of binary
pulsars in general, see Bhattacharya \& van~den~Heuvel (1991).

Millisecond pulsars in the NSWD systems are characterized by short 
rotational periods ($P_{\rm spin}<90$ ms) and relatively weak surface
magnetic fields ($B<10^{10}$ G) and are therefore fairly easy to
distinguish observationally from young non-recycled pulsars expected
to be found in the WDNS systems. Furthermore, one expects the NSWD systems
to have circular orbits ($e\ll 0.1$) since tidal forces
will have circularized the orbit very efficiently in the final
mass-transfer process (Verbunt \& Phinney 1995) 
-- as opposed to the high eccentricities
expected in systems where the last formed degenerate star
is a neutron star born in a supernova explosion (Hills 1983).\\
However, it is difficult to
determine from radio pulses alone whether an observed non-recycled 
pulsar belongs to a WDNS system, or if it is the last formed neutron star
in a NSNS system. The reason for this is the unknown inclination
angle of the binary orbital plane, with respect to the line-of-sight, which 
allows for a wide range of possible solutions for the mass of
the unseen companion. 
This is why PSR~B2303+46 was considered as a candidate for a
double neutron star system since its discovery (Stokes~et~al. 1985).
Only an optical identification of the companion
star could determine its true nature (van~Kerkwijk \& Kulkarni 1999).
The very recently discovered non-recycled binary pulsar
PSR~J1141--6545 (Manchester~et~al.~2000) has a massive companion
in an eccentric orbit ($M_2 \ge 1.0\,M_{\odot}$ and $e=0.17$).
This system must therefore be a new candidate for a WDNS system.
Only a deep optical observation will be able to
distinguish it from the NSNS alternative.

The identification of PSR~B2303+46 as a WDNS binary is very important for
understanding binary evolution, even though its existence 
already had been predicted by most binary population synthesis codes
(e.g. Tutukov \& Yungelson 1993).
Below we will demonstrate how to form a WDNS system and put
constraints on the initial parameters of the progenitor
of PSR~B2303+46. It should be noted that
the formation of PSR~B2303+46 has also recently been discussed by
Portegies~Zwart \& Yungelson (1999) and Brown~et~al. (2000).

In Sect.~2 we briefly introduce our evolutionary code,
and in Sect.~3 we outline the formation of a WDNS binary.
In Sect.~4 we discuss our results and in Sect.~5 we compare with
observations of PSR~B2303+46 and PSR~J1141--6545. 
The conclusions are summarized in Sect.~6.

\section{A brief introduction to the population synthesis code}
Monte Carlo simulations on a large ensemble of binary systems enables
one to examine the expected characteristics of a given binary
pulsar population and the physics behind the interactions
during their evolution.
We have used an updated version of the numerical population synthesis
code used by Tauris \& Bailes (1996). 
This code follows the evolution of a binary system
from the zero-age main sequence (ZAMS) to its ``final'' state, at all
stages keeping careful track of the mass and orbital separation
of the two stars. 
A large number of outcomes is possible from massive
binary evolution (see Dewey \& Cordes 1987) ranging from
systems which merged in a common envelope or became disrupted 
at the time of the supernova explosion, to binary pulsars with white dwarf,
neutron star or black hole companions. In these computations we restrict our
attention to systems which are likely to form WDNS binaries.

To simulate the formation of WDNS binaries, we assume that the initial 
system consists of two ZAMS stars in a circular orbit.
In our code we used a flat logarithmic initial separation distribution
($\Gamma (a) \propto a^{-1}$) and assumed a Salpeter initial mass function
for the ZAMS primary stellar masses
of $N(m) \propto m^{-2.35}$ combined with a mass-ratio function:
$f(q) = 2$/$(1+q)^2$ (Kuiper 1935). We adopt the term ``primary''
to refer to the {\it initially} more massive star, regardless of
the effects of mass transfer or loss as the system evolves.

In our simulations we used interpolations of the evolutionary grids
of Maeder \& Meynet (1988, 1989) for initial masses below 12 $M_{\odot}$.
These grids take into account stellar-wind mass loss in massive stars
and giant-branch stars, and also include a moderate amount of
overshooting from the convective core. For helium stars, we used the
calculations of Paczynski (1971).
From these models of stellar evolution we estimate the radius and age
of the primary star at the onset of the mass transfer.
Using other models with different chemical composition and
convective overshooting may have changed these values slightly,
but would not have had any significant effect on the parameters of
the final population of WDNS systems.
We refer to Tauris \& Bailes (1996) and Tauris~(1996) for a more
detailed description of the binary interactions and the computer code.

\section{Formation of a white dwarf -- neutron star binary}
In Fig.~1 we depict the scenario for making a WDNS system
and in Fig.~2 we have shown the masses of the initial ZAMS stars
which evolve to form WDNS systems. The allowed parameter space is
constrained by 4 boundaries described below. 
\begin{figure}
  \psfig{file=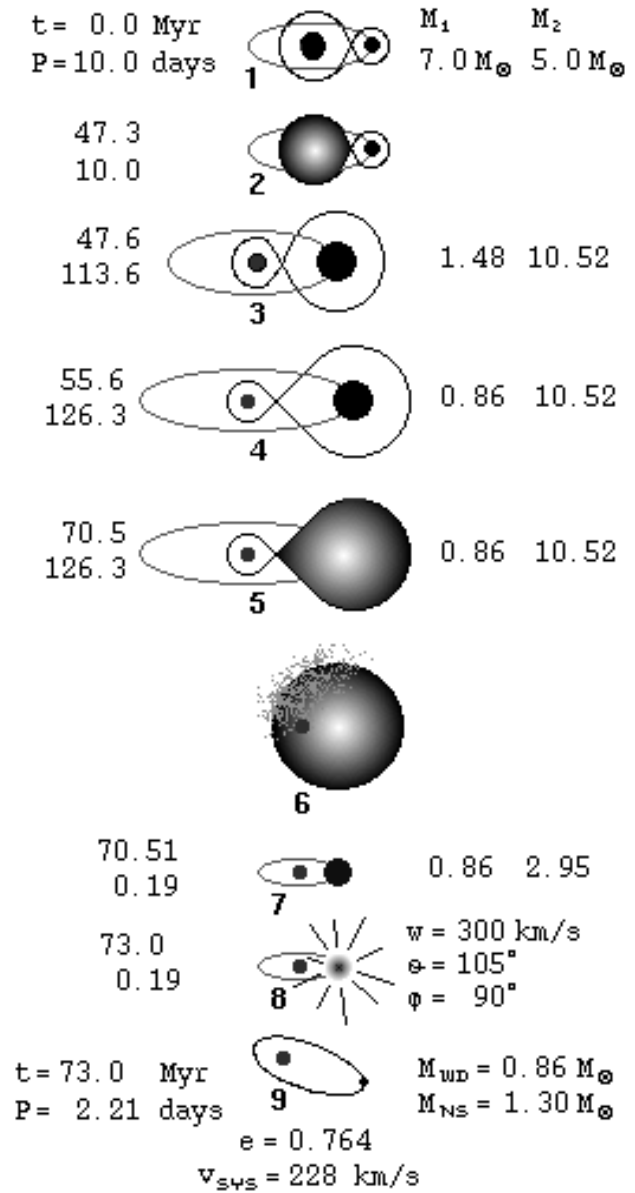,width=\columnwidth}
     \caption{Rough illustration of the formation of a typical WDNS system.
              The neutron star is formed last and is therefore
              not recycled
              -- see text.}
\end{figure}
In order to successfully form a WDNS system we found it necessary
to choose primary masses, $M_1$ in the interval $5-11\,M_{\odot}$,
secondary masses, $M_2$ between $3-11\,M_{\odot}$, and 
initial separations, $a_0$ between $5-600\,R_{\odot}$.
In order for the primary to
end its life as a white dwarf we must require $M_1<11\,M_{\odot}$
(cf. boundary {I} in Fig.~2).
On the other hand a minimum mass of $\sim 5\,M_{\odot}$ is needed
for the primary, since it has to transfer sufficient mass
to the secondary (which is initially lighter than the primary)
in order for the latter to explode in a supernova (SN) once its nuclear
burning has ceased. The requirement of substantial mass accretion
onto the secondary means that the progenitor binary can not evolve
through a common envelope (CE) in the first phase of mass transfer
(cf. stage~2 in Fig.~1).
The reason is that the timescale for the CE-phase is very
short ($10^3 - 10^4$~yr) compared to the duration of a dynamically stable
Roche-lobe overflow (RLO) which lasts for $\sim 1$ Myr.
In such a short time hardly any mass can be accreted.
We assumed in our calculations the RLO to be conservative ({\em i.e.} 
total mass and orbital angular momentum remains constant).
In order to avoid a CE-phase we required $q=M_2/M_1>0.4$
(cf. boundary {II} in Fig.~2).
Hence $M_2$ is constrained from this requirement in
combination with: $M_2+\Delta M_{\rm RLO} > M_{\rm SN}^{\rm crit}$,
where $M_{\rm SN}^{\rm crit}$ is the threshold mass for undergoing
a supernova explosion and $\Delta M_{\rm RLO}$ is the amount
of matter accumulated by the secondary from the primary star
during the RLO, cf. boundary {III} in Fig.~2.
$M_{\rm SN}^{\rm crit}$ is typically $9-11\,M_{\odot}$
but depends on its core mass, $M_{\rm He}^{\rm crit}$
and its evolutionary status at the onset of the mass transfer
(e.g. Bhattacharya \& van~den~Heuvel 1991).
It should therefore be noted that the boundaries in Fig.~2
depend on $P_{\rm orb}$ since the evolutionary
status of a donor star is important in order to determine
$M_{\rm He}^{\rm crit}$. Also the onset criterion for the
formation of a CE depends on the evolutionary status of the
Roche-lobe filling star (see Sect.~5.3.1 for further discussion). 
The last boundary ({IV}) in Fig.~2 results from
the trivial requirement that the secondary star
initially has $M_2<M_1$.

After the primary star has lost its envelope to the secondary,
the helium core will eventually settle as a white dwarf\footnote 
{It should be noted here, that we have neglected the expansion of the
naked helium star which might otherwise have caused a 
subsequent rapid mass-transfer phase (cf. discussion in Sect. 5.3.2).}
(stage~3 and 4, respectively in Fig.~1). To calculate the final mass
of the white dwarf we assumed:
\begin{equation}
  M_{\rm WD}=\left\{ \begin{array}{ll}
      M_{\rm 1He} & \mbox{\hspace{0.5cm}$M_{\rm 1He}<0.45\,M_{\odot}$}\\
      0.27\,M_{\odot}+0.40\,M_{\rm 1He} & 
                   \mbox{\hspace{0.5cm}$M_{\rm 1He}>0.45\,M_{\odot}$}\\
                      \end{array}
         \right.
\end{equation}
(unless $M_{\rm 1He}>M_{\rm He}^{\rm crit}$
in which case the mass of primary helium core is above the threshold limit
$\sim 2.5\,M_{\odot}$ and it collapses in a supernova 
leaving behind a neutron star remnant). 
All the white dwarfs formed in WDNS binaries will be a {CO} or {O}-{Ne}-{Mg}
white dwarf with mass $M_{\rm WD}>0.60\,M_{\odot}$.
Beware that the above ad hoc
formula does not exactly represent the observed distribution of binary
white dwarf masses which depend on the orbital period as well.

The subsequent evolution is now reversed.
The secondary fills its Roche-lobe and initiates mass transfer
onto the white dwarf (stage~5 in Fig.~1). This class of binaries
is referred to as symbiotic stars in the literature
(a sub-class of cataclysmic variables).
Now because of the extreme mass
ratio between the secondary star (the donor) and the accreting
white dwarf, the orbit will shrink upon mass transfer.
Furthermore, if the secondary star is a red giant at this stage
(which is almost always the case) it has a deep convective envelope
and will expand in response to mass loss. This enhances the
mass-transfer rate, which in turn causes the orbit to shrink
faster. This leads to a run-away event and the transferred
material piles up and is heated in a cloud around the
white dwarf. 
The white dwarf will be embedded in a CE with the secondary star
(stage~6 in Fig.~1).
In this dynamically unstable situation drag forces causes the
white dwarf to spiral-in toward the core of the giant companion
(e.g. Iben \& Livio 1993).
In this process orbital energy is converted into kinetic energy
which provides outward motion and possibly ejection of the envelope.
Here we assumed an efficiency parameter of $\eta _{\rm CE}=2.0$.
If there is not enough orbital energy
available to expel the envelope (i.e. the binary
is too tight), the result is that the white dwarf will merge
with the core of the giant -- perhaps leading to a type~Ia SN
and total disruption of the binary (in any case we terminated
our calculations for binaries with merging stars). 
This puts some constraints
on the orbital separation at the moment of mass transfer.
If the envelope is ejected successfully (stage~7), the helium core
will collapse after a short while and explode in a type Ib/c supernova
(stage~8). Also here we have ignored the expansion of the helium star
as a result of large uncertainties in the orbital evolution caused
by rapid mass transfer of 0.3 -- 0.5$\,M_{\odot}$ from the helium star
-- cf. Sect. 5.3.2.\\
There is plenty of evidence from observations
(e.g. Lyne \& Lorimer 1994; Tauris~et~al.~1999) that a
momentum kick is imparted to newborn neutron stars.
This is probably due to some asymmetry in the neutrino emission
driving the explosion. We assumed an isotropic distribution of kick 
directions and used a Gaussian distribution for its magnitude
with a mean value, $\langle w \rangle = 500$~km~s$^{-1}$ and
standard deviation, $\sigma = 200$~km~s$^{-1}$.
We assumed the neutron star to be born with a mass of
$M_{\rm NS}=1.3\,M_{\odot}$. Given the pre-SN separation and the mass
of the collapsing star and its white dwarf companion,
as well as the kick magnitude and direction, we are
able to calculate the final orbital period, $P_{\rm orb}$
and eccentricity, $e$ of the WDNS system (cf. stage~9 in Fig.~1).
\begin{figure}
  \psfig{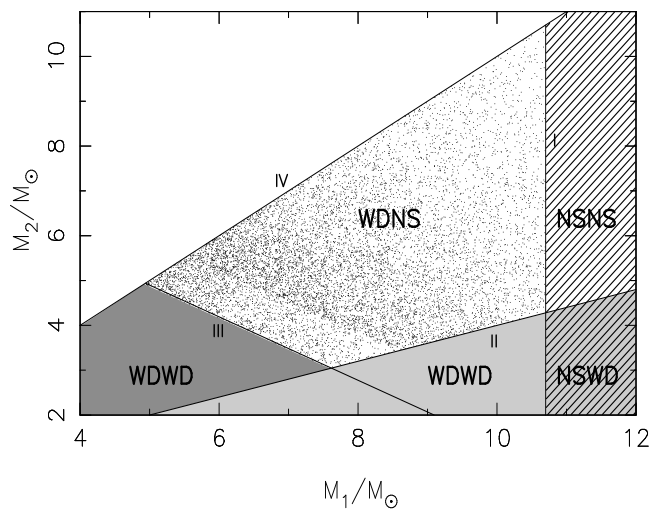}
     \caption{Area in the ZAMS ($M_1,M_2$) plane with 
      progenitor systems for WDNS binaries -- see text for a discussion.}
\end{figure}

\section{Results}
\begin{figure*}
  \psfig{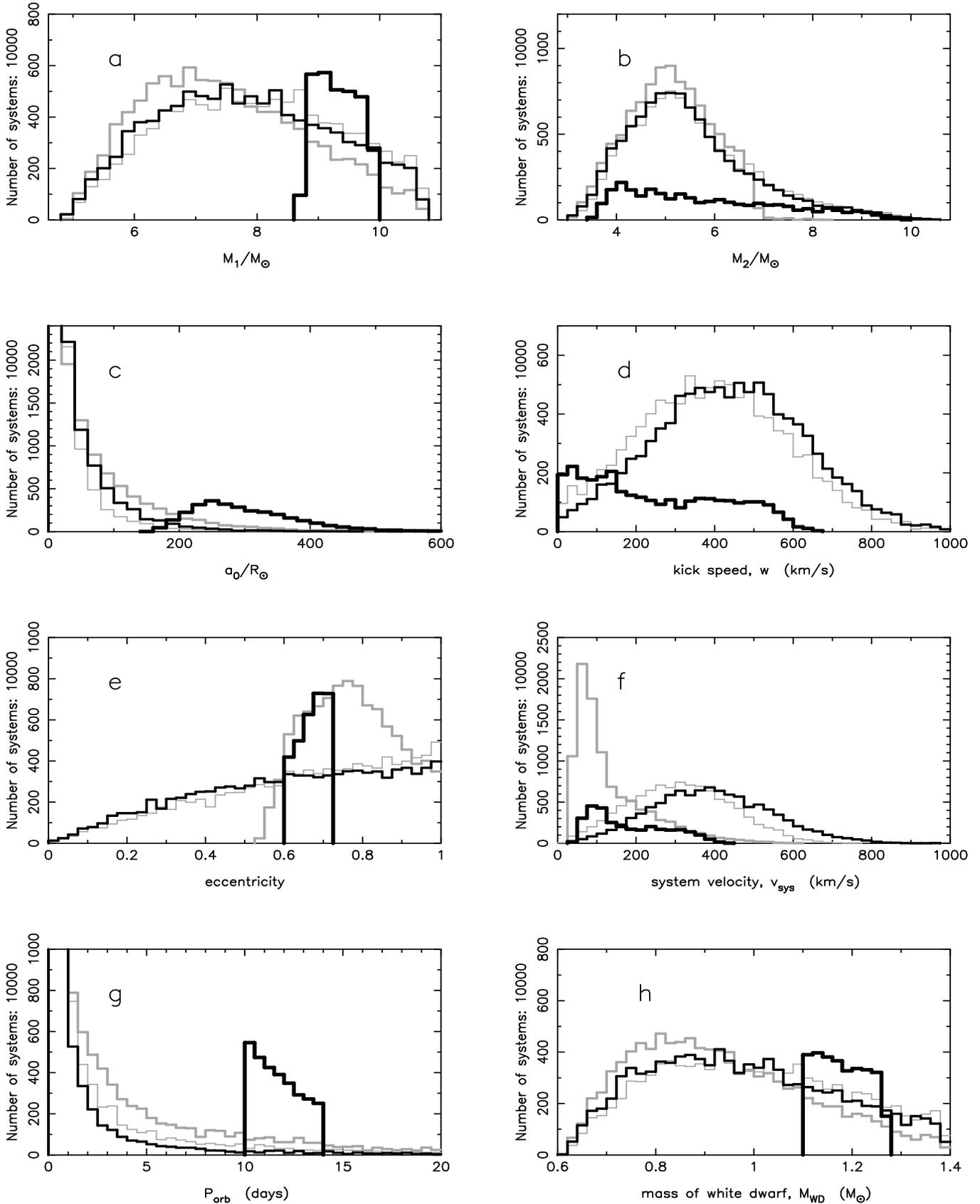}
     \caption{Distributions of parameters for the initial
              ZAMS systems (panels a--c) which evolved to
              form WDNS binaries. The final parameters
              of these WDNS systems are plotted in 
              panels e--h. See text for further explanation of
              the different curves.}
\end{figure*}
The results of our computations are summarized in Fig.~3
which shows distributions of the parameters of both
the ZAMS progenitor binaries (panels a--c) and the observable
parameters of the final WDNS binaries (panels e--h).
Each panel shows four different distributions of a given parameter
for 10$\,$000 systems which successfully evolved to a WDNS binary.
These distributions are represented by a thin black, gray, thin gray
and thick black histogram, respectively.

\subsection{The thin black curve: asymmetric SN}
The thin black curve represents our standard simulation where
we assumed $\langle w \rangle = 500$~km~s$^{-1}$, with
$\sigma = 200$~km~s$^{-1}$, for the magnitude of kicks
in the SN, and a CE efficiency parameter of $\eta _{\rm CE}=2.0$.
We now describe the distribution of parameters represented
by this thin black curve in each panel.

From panels a--c we conclude that in order to form a WDNS binary
one must have a ZAMS binary with $4.9<M_1/M_{\odot}<10.8$,
$3.0<M_2/M_{\odot}<10.8$ and $a_0 < 600\,R_{\odot}$.
We note that the distribution of kicks in panel~d closely follows our
assumed input distribution (which is somewhat unknown ad hoc
-- see e.g. Hartman~et~al.~1997 for a discussion). 
It is also seen that
the distribution of kicks imparted in binaries which form bound
WDNS systems is slightly shifted towards smaller values compared
to the trial distribution (i.e. on average the WDNS systems
have received a kick of 439~km~s$^{-1} \neq 500$~km~s$^{-1}$).
The reason is simply that, on average, a system has a higher
probability of surviving the SN if the kick is small.
The maximum kick possible in the SN appears to be 1000~km~s$^{-1}$.
However, this is an artifact from our trial distribution
of kicks. Applying unlimited values for the magnitude of the kicks
it is in principle possible to form a WDNS system using 
a kick, $w > 1700$~km~s$^{-1}$ --
resulting in space velocities of $v_{\rm sys} > 1300$~km~s$^{-1}$.\\
The expected orbital parameters of the formed WDNS systems
are shown in panels e--h. The WDNS systems can be formed with
any eccentricity, $e$ -- though the distribution increases
monotonically with increasing values of $e$. The simulated
systemic 3-D velocities of the WDNS systems (panel~f) are simply correlated
to the distribution of kicks.
When comparing with future observations of 2-D transverse
velocities one has to multiply this simulated 3-D distribution 
by a factor of $\pi/4$ in order to separate out the unknown radial velocities.\\
The distribution of orbital periods, $P_{\rm orb}$ of the WDNS binaries
is highly concentrated towards small values of $P_{\rm orb}$
(notice we have not shown the first two bins of the distribution
in panel~g, which reach values of 6500 and 1100, respectively).
This is also seen in Fig.~4 where we have
plotted the eccentricity versus the final orbital
period for the simulated WDNS systems.
The cumulative probability curve for $P_{\rm orb}$ is also shown in
the figure.
The distribution of calculated white dwarf masses (in panel~h of Fig.~3)
is seen to be in the interval $0.6<M_{\rm WD}/M_{\odot}<1.4$.

We will now briefly describe what happens when we choose other
input distributions governing the two most important binary
interactions: the SN and the preceding CE-phase.

\subsection{The gray curve: symmetric SN}
The gray curve in each panel shows the distribution 
of a given parameter assuming a purely
symmetric SN. This assumption apparently does not 
significantly change the distributions of initial parameters 
for the binaries ending up as WDNS systems. It has only a
moderate effect on the distribution of $M_2$ and $M_{\rm WD}$
(leading to slightly lower values),
but a more significant effect on $a_0$ and the distribution of
final $P_{\rm orb}$, which tend to be higher without the
asymmetry in the SN. This is expected, since in the cases where
a kick is added on top of the effect of sudden mass loss, the
pre-SN orbit must, on average, be more tight in order to survive the SN.
However, most important differences are seen in the
distributions of the eccentricity and the systemic velocity.
The shift to smaller values for the latter distribution
(the mean value of $v_{\rm sys}$ decreases from 393 
to 137~km~s$^{-1}$ with no kicks) follows naturally, since now $w=0$ and
$v_{\rm sys}$ is therefore only arising due to the
sudden mass loss in the SN.\\
In the case of post-SN eccentricities, we now only
form WDNS systems with $e>0.5$. The existence of a minimum
eccentricity for systems undergoing a symmetric SN
follows from celestial mechanics (e.g. Flannery \& van~den~Heuvel 1975):
\begin{equation}
  e = \frac{M_{\rm 2He}-M_{\rm NS}}{M_{\rm WD}+M_{\rm NS}}
\end{equation}
Since we have assumed
$M_{\rm 2He}>M_{\rm He}^{\rm crit}\simeq 2.5\,M_{\odot}$
and $M_{\rm WD}^{\rm max}< 1.4\,M_{\odot}$ it follows
that $e>0.45$.

Another proof of asymmetric SN in nature would therefore 
be a future detection of a WDNS system with $e<0.45$ 
(in Sect.~5.3.2 we discuss the effects of 
a possible case BB RLO from the helium star prior to the SN,
in which case the minimum eccentricity for symmetric SN
becomes smaller).
The reason why asymmetric SN can result in more circular
orbits is a result of the expected randomness in the direction
of the kicks. Hence a system which would otherwise have been
highly eccentric after a symmetric SN (i.e. due to the
effect of sudden mass loss alone) can end up being almost
circular when momentum is imparted to the newborn neutron star
in a particular direction.\\
It should be noted that the post-SN orbits of the WDNS systems
will not circularize later on, since tidal effects are
insignificant in systems with two degenerate stars.
Only in binaries with $P_{\rm orb} \la 0.6$ days will the
orbit circularize in less than a Hubble-time -- mainly as a result
of general relativistic effects (Shapiro \& Teukolsky 1983).

\subsection{The thin gray curve: high efficiency in ejection of the CE}
The thin gray line in each panel shows a given distribution 
assuming $\eta_{\rm CE}=8.0$. This is a radical change
in the efficiency of ejecting the envelope.
The efficiency of converting
orbital energy into kinetic energy, which expels the envelope, is defined by:
$\Delta E_{\rm bind} \equiv \eta _{\rm ce} \, \Delta E_{\rm orb}$ (Webbink 1984), or:
\begin{equation}
  \frac {GM_2M_2^{\rm env}}
   {\lambda a_{\rm i} r_{\rm L}}
  = \eta _{\rm ce}\left[ \frac {GM_{\rm WD}M_{\rm 2He}}{2a_{\rm f}} -
  \frac {GM_{\rm WD}M_2}{2a_{\rm i}} \right]
\end{equation}
which yields the following change in orbital separation:
\begin{equation}
  \frac {a_{\rm f}}{a_{\rm i}} = \frac {M_{\rm WD}M_{\rm 2He}}{M_2}
  \frac {1}{M_{\rm WD} + 2M_{\rm 2env}/
  (\eta _{\rm ce}\lambda r_{\rm L})}
\end{equation}
where $r_{\rm L}$ = $R_{\rm L}/a_{\rm i}$ is the dimensionless Roche-lobe radius
of the donor star, $\lambda$ is a weighting factor ($\le $ 1.0) for
the binding
energy of the core and envelope of the donor star and, finally,
where $M_{\rm 2He},
M_{\rm 2env},a_{\rm i}$ and $a_{\rm f}$ are the mass of
the core and hydrogen-rich envelope of the evolved companion star,
and the pre-CE and post-CE separation, respectively.
We notice that this large change in $\eta_{\rm CE}$ barely changes our results.
The thin gray curve seems to follow the thin black curve closely in most of the panels.
Only the final $P_{\rm orb}$ are higher in this case --  though the
shape of the distribution remains the same.
This result is expected since in this case the envelope
is ejected more easily in the CE, resulting in larger pre-SN separations.
One can easily show that roughly speaking: 
$(a_{\rm f}/a_{\rm i}) \propto \eta _{\rm CE}$ and hence we expect
the pre-SN binaries to have much wider orbits
in the case where $\eta _{\rm CE}=8.0$ (by a factor of 4).
However, many of the wide pre-SN binaries do not survive
the SN, so the overall final $P_{\rm orb}$ distribution only differs
slightly from the case with $\eta _{\rm CE}=2.0$.
This also explains why the selected $w$ and resulting
$v_{\rm sys}$ are slightly lower,
since binaries are more easily disrupted having wider orbits.
Thus $w$ and $v_{\rm sys}$ of the binaries which do survive the SN
are smaller in this case where $\eta _{\rm CE}=8.0$.

We also experimented with different initial mass functions (IMF)
but conclude that our simulated results for the WDNS systems are quite robust
against changes in the IMF, the mass-ration function $f(q)$
or the initial separation function for $a_0$ (see below).

\section{Comparison with observations}
The observed parameters of the binary radio pulsars PSR~B2303+46 and PSR J1141--6545
are given in Table~1.

\subsection{PSR B2303+46}
Knowing the observed parameters for the orbital period, eccentricity
and mass function, $f$ we have investigated which initial ZAMS
binaries would form a system like PSR~B2303+46 and also constrained
its systemic velocity. 
\begin{figure}[t]
  \psfig{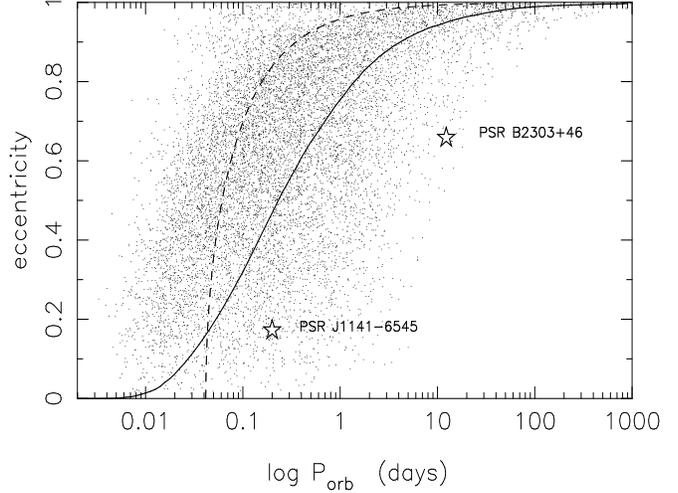}
     \caption{Eccentricity versus final orbital period 
for the simulated WDNS systems. The curve gives the probability that
$P_{\rm orb}$ for a simulated WDNS system is less than the corresponding
value on the x-axis. We assumed here $\langle w \rangle = 500$~km~s$^{-1}$,
$\sigma = 200$~km~s$^{-1}$ and $\eta _{\rm CE}=2.0$. Systems formed to
the left of the dashed line will coalesce within $\sim 10$~Myr as a result
of gravitational wave radiation -- see text.}
\end{figure}
The distribution of parameters for 3000 WDNS systems which resemble PSR~B2303+46
are shown as the thick black histograms in Fig.~3.
We define a WDNS system to roughly resemble PSR~B2303+46 if $10<P_{\rm orb}<14$ days,
$0.60 < e < 0.72$ and $M_{\rm WD}>1.1\,M_{\odot}$.
The minimum white dwarf mass given in Table~1 is obtained from its
observed mass function. However a tighter constraint is found in
combination with the 
measurement of the general relativistic rate of periastron advance
(Thorsett~et~al.~1993; Arzoumanian 1995) which
yields a solution for the total system mass:
$M_{\rm NS}+M_{\rm WD}=2.53 \pm 0.08\,M_{\odot}$, and hence
$M_{\rm WD}>1.20\,M_{\odot}$.
In order to produce such a high mass for the white dwarf we need a minimum
mass $M_1>8.8\,M_{\odot}$ for its progenitor, but notice that the
wide range of possible secondary masses remains.

The most distinct feature of PSR~B2303+46 is its very wide orbit: 
$P_{\rm orb}=12.3$ days. It is evident from Fig.~4, and panel~g in Fig.~3,
that such a large value for $P_{\rm orb}$ is very rare.
We find that only $\sim 5 \%$ of all WDNS systems are formed with
$P_{\rm orb} \ge 12.3$ days (see the cumulative probability curve
in Fig.~4). The large observed value of $P_{\rm orb}$ leads to a large initial 
separation of the ZAMS binaries in the interval: $160 < a_0/R_{\odot} < 600$.
To test whether or not the resulting $P_{\rm orb}$ distribution for the
WDNS systems simply reflects our initial trial distribution for $a_0$
(which is flat in $\log a_0$), we tried to run our code with a constant
distribution function (i.e. all values of $a_0$ are a priori
equally probable). This however, does not change our conclusions and
we infer that the distribution of $P_{\rm orb}$ is highly concentrated
toward small values.

For the expected value of the recoil velocity of PSR~B2303+46
we find it must be in the interval: $50 < v_{\rm sys} < 440$~km~s$^{-1}$,
with a mean value of 181~km~s$^{-1}$. 

\begin{table}[t]
\setlength{\tabcolsep}{5.0pt}  
  \caption{Observed parameters of the WDNS system PSR B2303+46 and 
           the WDNS candidate PSR J1141--6545. 
           We use $M_{\rm NS}=1.3\,M_{\odot}$.}
  \begin{center}
   \begin{tabular}{lllccrc}
   \noalign{\smallskip}
        \hline
        \hline\noalign{\smallskip}
    PSR$\qquad$ & $P_{\rm orb}$  & f & $M_{\rm WD}^{\rm min}$ & 
         ecc. & $P_{\rm spin}$ & $\dot{P}_{\rm spin}$\\
         \noalign{\smallskip}
         \hline\noalign{\smallskip}
 B2303+46    & 12.3 & 0.2464 & 1.13  
     & 0.658 & 1066 & 0.57\\
 J1141--6545 & 0.198 & 0.177  & 0.97  
     & 0.172 &  394 & 4.31\\
         \hline\noalign{\smallskip}
                & days & $M_{\odot}$ & $M_{\odot}$ & 
              & ms & 10$^{-15}$\\
          \noalign{\smallskip}
          \hline
          \hline
     \end{tabular}
  \end{center}
\end{table}
\subsection{PSR J1141--6545}
This interesting binary radio pulsar was recently found in the ongoing
Parkes Multibeam Survey (Manchester~et~al. 2000).
The high value of $\dot{P}_{\rm spin}$
in combination with its relatively slow rotation rate 
($P_{\rm spin}=394$ ms)
and non-circular orbit ($e=0.17$) identifies this pulsar as being
young and the last formed member of a double degenerate system.
Given $P_{\rm orb}=0.198$ days it is evident that a non-degenerate
star can not fit into the orbit without filling its Roche-lobe.
The minimum companion mass of $0.97\,M_{\odot}$ 
(assuming $M_{\rm NS}=1.3\,M_{\odot}$) and its location in the
($P_{\rm orb},e$) diagram makes this binary a good candidate
for a WDNS system. We note, that $P_{\rm orb}=0.198$ days  
is right at the mean value of the simulated distribution --
though the eccentricity is lower than average for this
orbital period. We must however, beware of selection effects
at work here. It is well known that it is much more difficult
to detect a periodic signal emitted from an accelerated pulsar 
in a narrow orbit compared to a wide orbit 
(e.g. Ramachandran \& Portegies~Zwart 1999).
It is therefore expected that future detections of WDNS
systems will fill up mainly the area to the right side of the curve 
in Fig.~4. Although some WDNS systems should be found in very
tight orbits from improved future acceleration searches, 
it should also be noted that tight binaries cannot avoid merging
as a result of emission of gravitational waves 
(Landau \& Lifshitz 1958; Peters 1964).
Given the values of $P_{\rm orb}$ and $M_{\rm WD}^{\rm min}$,
PSR~J1141--6545 will merge within $\sim 590$ Myr 
(a spectacular event for LIGO).
However, the average lifetime of a non-recycled pulsar is only
$\sim 10$ Myr (Taylor~et~al. 1993). In Fig.~4 we have shown
with a dashed line WDNS systems which will merge on a timescale
of $\sim 10$ Myr, assuming $M_{\rm NS}=1.3\,M_{\odot}$ and 
$M_{\rm WD}=1.0\,M_{\odot}$. Hence binaries found to the left of this line 
($P_{\rm orb} \la$ a few hr) are most likely to coalesce within the lifetime
of the observed radio pulsar.

As shown in Sect.~4.2 we conclude that the SN forming the observed pulsar
in PSR~J1141--6545 must have been asymmetric, given the low eccentricity
of the system. We find a minimum kick of 100~km~s$^{-1}$ necessary
to reproduce $e=0.17$ for a system with similar $P_{\rm orb}$.
More interesting, we are able to constrain the minimum space
velocity of PSR~J1141--6545, and we find: $v_{\rm sys} > 150$~km~s$^{-1}$.
A future proper motion determination and a possible detection of a spectral line
from the white dwarf will be able to verify this result.

\subsection{On the relative formation rates of WDNS, NSNS and NSWD systems}
We have used our population synthesis code to generate a large number of 
compact binaries using wide initial trial distributions for the ZAMS binaries.
In Table~2 we show the relative formation rates of these systems (normalized
to the number of trial simulations). 
The remaining binaries either evolved into WDWD systems, coalesced,
became disrupted in a SN or formed a black hole binary.
It should be noted that many of the parameters governing the binary
interactions are not very well known, and that the relative formation
rates are sensitive to these parameters.
However, the aim of this little exercise is to demonstrate it
is possible to match the present observations with
the results of simple evolutionary simulations.\\
To estimate the relative Galactic abundancies of {\em active}
WDNS, NSNS and NSWD binaries, we have weighted each
of the formation rates by the observable 
lifetime of the pulsars, $\tau _{\rm psr}$.\\
For non-recycled pulsars we chose a lifetime of 10 Myr. For the recycled
pulsars $\tau _{\rm psr}$ depends on the mass-transfer timescale and
the amount of material accreted (and hence on the mass of the progenitor
of the last formed degenerate star). The mass transfer causing the recycling 
in the NSNS systems takes place on a sub-thermal
timescale. Hence the rejuvenated $\dot{P}$ is relatively large
($\sim 10^{-18}$) and these pulsars are expected to terminate their radio
emission in less than about one Gyr. Hence we chose 
$\tau _{\rm psr} = 1000$ Myr for the NSNS binaries with an observed
recycled pulsar.
The situation is similar for the NSWD systems with heavy {CO}/{O}-{Ne}-{Mg}
white dwarf companions. However, the binary NSWD systems with
low-mass helium white dwarf companions are expected to have
been recycling over a long interval of time ($\sim 100$ Myr)
and exhibit long spin-down ages of 5--10 Gyr.  
We therefore chose an overall average value of $\tau _{\rm psr}=3000$ Myr
for the entire NSWD population consisting of both systems with a {CO} or
a low-mass helium white dwarf.\\ 
For these values of $\tau _{\rm psr}$, in combination with our
simulated relative birthrates, we find the relative number
of such systems expected to be observed, $N_{\rm sim}$ (normalized to
the total actual number of binary pulsars detected so far).
We see that our simple estimates matches well with
the observations -- cf. $N_{\rm sim}$ and $N_{\rm obs}$ in
columns 4 and 5, respectively in Table~2.

The actual values of $\tau _{\rm psr}$ are uncertain,
and a weighting factor should also be introduced to correct for
the number of different systems merging from gravitational wave radiation. 
Furthermore, when comparing with observations
one must also correct for other important effects, such as the
different beaming factors (i.e. the area on the celestial
sphere illuminated by the pulsar) between recycled and non-recycled
pulsars. Slow non-recycled pulsars usually have much more narrow
emission beams than those of the recycled pulsars and hence
they are more unlikely to be observed. On the other hand their
pulse profiles are less likely to be smeared out.

We find the {\em birth} rate of WDNS systems to be considerably
higher than that of the NSNS systems (by more than an order
of magnitude).
In a recent paper Portegies~Zwart \& Yungelson (1999) also
investigated the formation of WDNS binaries. They find a roughly
equal birth rate of WDNS and NSNS systems. The reason for the
discrepancy with our results is mainly that they assume the initial 
mass-transfer process (stage 2 in Fig.~1) to be non-conservative
so the majority of the transfered matter is lost from the system.
Therefore much fewer secondary stars in their scenario accrete
enough material to collapse to a neutron star later in the evolution.
\begin{table}[t]
\setlength{\tabcolsep}{6.0pt}  
  \caption{Estimated relative formation rates as well as simulated
           and observed Galactic disk abundances  
           of WDNS, NSNS and NSWD binaries. The observed globular cluster
           pulsars are not included since they are formed via exchange
           collisions in a dense environment.}
  \begin{center}
   \begin{tabular}{rcrcr}
   \noalign{\smallskip}
        \hline
        \hline\noalign{\smallskip}
     binary & rel. birthrate & $\tau _{\rm psr}$ & $N_{\rm sim}$ & $N_{\rm obs}$\\
         \noalign{\smallskip}
         \hline\noalign{\smallskip}
        WDNS & 0.00567 & 10 Myr & 1.37 & 2\\
        NSNS & 0.00032 & 10 Myr & 0.08 & 0\\
        ---  & ---  & 1000 Myr & 7.75 & 5\\
        NSWD & 0.00052 & 3000 Myr & 37.8 & 40\\
          \noalign{\smallskip}
          \hline
          \hline
     \end{tabular}
  \end{center}
\end{table}

\subsubsection{The onset criteria of a common envelope}
In this work we assumed all binaries will avoid evolving into
a CE during the initial mass-trasnfer phase (stage 2) if $q>0.4$.
This is, of course, only a rough boundary condition which 
requires a much more careful investigation of the evolutionary
status of a given binary followed by detailed hydrodynamical calculations
of the mass-transfer process (Iben \& Livio 1993).
We also simulated the formation of WSNS systems using a limiting
value of $q<0.7$ for the formation a CE. In this case we produce
relatively more NSWD systems (by a factor of $\sim 2$) and
fewer WDNS and NSNS (also by a factor $\sim 2$ in each case).
This can be seen by moving up the boundary {II} in Fig.~2.
However, it is important that the birth ratio of WDNS to NSNS systems
remains about the same ($\sim$ 18:1).

\subsubsection{The evolution of naked helium stars}
The outer radii of low-mass helium stars ($M_{\rm He}<3.5\,M_{\odot}$)
may become very large during the late evolutionary stages
(Paczynski 1971; Habets 1985) and could initiate a second phase
of mass transfer (so-called case BB mass transfer, Delgado \& Thomas 1981).\\
If $M_{\rm He} \ga 2.5\,M_{\odot}$ the helium stars will develop a core mass
larger than the Chandrasekhar limit at the onset of central
or off-center convective core-carbon burning. 
These stars are therefore expected to explode in a SN and leave
neutron star remnants. However, the masses of such collapsing stars
may have been reduced by some 10--20$\,$\% as a result
of mass transfer in a close binary and a strong stellar wind
when the helium stars expanded. Notice, that helium stars only lose
a fraction (30--50$\,$\%) of the envelope mass 
outside their carbon cores in case BB RLO, as opposed to normal giants
which usually lose their entire (hydrogen) envelopes in RLO.
Therefore, if the minimum mass of a collapsing star is assumed
to be only $2.0\,M_{\odot}$ the resulting minimum post-SN eccentricity will be 
0.26 (cf. Eq.~2) in the case of a symmetric SN
-- instead of 0.45 in the case of a $2.5\,M_{\odot}$ exploding star.
However, this value is still larger than $e=0.172$ observed in
PSR~J1141--6545 which indicates a kick was involved in its formation.
Even if the mass of the collapsing star would be as low as $1.8\,M_{\odot}$
a kick would still be necessary.

In our simple scenario of making WDNS binaries the effect of case BB
mass transfer would alter our calculations in stages 3 and 7 in Fig.~1.
However the effects on the global outcome are probably small. 
For a second mass transfer in stage 3, the specific angular momentum 
of matter lost from the system is poorly known and may result in either
shortening or widening of the orbit. Furthermore, the orbital
evolution is later dominated by the much more important spiral-in phase
in stage~6. If case BB RLO occurs in stage 7, mass transfer
from the helium star to
the (lighter) white dwarf decreases the orbit; but mass loss from the vicinity
of the white dwarf (due to extreme high mass-transfer rates),
or in the form of a direct stellar wind from the helium star,
will widen the orbit. Therefore the net effects are uncertain
and we believe our simple calculations presented here remain 
approximately valid.

There are many theoretical and observational selection effects involved 
in our simple work presented here
and we have a more rigorous analysis in preparation.

Finally, it should be mentioned that it may be possible (Nomoto 1982) that 
helium stars $1.4 \la M_{\rm He}/M_{\odot} \la 1.9$
ignite carbon in a violent flash (degenerate core) which might lead to total
disruption of the star (type Ia supernova).
If this picture is correct for a significant fraction of helium stars,
our estimate for the production rate of WDNS is overestimated.

\subsection{On the nature of PSR~J1141--6545} 
Despite various uncertainties described above one important
result can be extracted from our calculations: the
expected formation rate of WDNS systems is $\sim \,$18 times higher
than that of the NSNS systems (8--9 times higher if we require
$M_{\rm WD}>0.97\,M_{\odot}$, the minimum companion mass
observed in PSR~J1141--6545). The main reason is that the
WDNS binaries only need to survive one SN, and that their
progenitors are favored by the IMF.
Based on our simulations
we therefore conclude that since the pulsar in PSR J1141--6545
is young (non-recycled) it is most likely (at the $\sim \,$90$\,$\%
confidence level) to have a white dwarf companion.
This prediction will hopefully soon be verified or falsified
from optical observations.

\section{Conclusions}
\begin{itemize}
\item We have adapted a simple numerical computer code to study
      the formation process of WDNS systems and have demonstrated
      that many initial ZAMS binary configurations can result
      in the formation of such systems.
\item We have constrained the parameters 
      for the progenitor system of PSR B2303+46.
\item We have presented evidence in favor of a
      white dwarf companion (rather than a neutron star)
      to the newly discovered binary pulsar PSR~J1141--6545.
\item We have also demonstrated that an applied kick was necessary in
      the formation scenario of this system.
\item Finally, we predict PSR~J1141--6545 to have a minimum
      3-D systemic velocity, $v_{\rm sys} > 150$~km~s$^{-1}$.
\end{itemize}

\acknowledgements{We thank Dick Manchester
and the Parkes Multibeam Survey team for releasing the
orbital parameters of PSR~J1141--6545 prior to publication. 
We thank Norbert~Wex, Ed~van~den~Heuvel and Lev~Yungelson
for comments on an earlier version of this manuscript.
We also thank the referee, Dr. M.~van~Kerkwijk, for his comments.
T.M.T. acknowledges the receipt of a Marie Curie Research Grant
from the European Commission.}

\end{document}